\documentstyle[seceq,epsf,wrapft]{ptptex}

\def\vS{{\mib S}}
\def\cH{{\cal H}}

\title{
Level Spectroscopy: Physical Meaning and Application to the
Magnetization Plateau Problems
}

\author{
Kiyomi {\sc Okamoto}\footnote{E-mail: kokamoto@stat.phys.titech.ac.jp}
}

\inst{
Department of Physics, Tokyo Institute of Technology, Tokyo 152-8551
}

\recdate{
~~~~~~~~~~~~~~~~~~~~~~~~
}

\abst{
We review the {\it level spectroscopy},
which is a powerful method of analyzing the numerical data with respect to
the Berezinskii-Kosterlitz-Thouless quantum phase transition
in one dimension.
We focus on its physical meaning and also its application to the magnetization
plateau problems.}

\begin{document}

\maketitle

\section{Introduction}
The Berezinskii-Kosterlitz-Thouless\cite{bere,KT,kosterlitz,kogut} (BKT, or simply KT)
quantum phase transition in one-dimensional systems
is the transition between the gapless state and the gapful state.
As is well known, the critical behavior of the BKT transition is
highly singular and sometimes called \lq\lq pathological\rq\rq.
In finite systems, this high singularity appears as severe
slowly-converging logarithmic size corrections in various physical quantities.
Thus it is very difficult to determine the BKT quantum phase transition point
from the numerical data if we use the conventional methods.

We note that there are two types in the BKT quantum phase transition.
One is the doubly degenerate type, of which examples are
(a) the transition between the spin-fluid (SF) and N\'eel states
in the $S=1/2$ $XXZ$ spin chain, and
(b) the SF-dimer transition in the $S=1/2$ Heisenberg spin chain
with next-nearest-neighbor interactions.\cite{tonegawa-harada,ON,NO1,NO2}
In these cases, the gapped states
(the N\'eel state of (a) and the dimer state of (b) are
doubly degenerate.
The mechanism of the transition of this type is the
spontaneous symmetry breaking.
Another type is the non-degenerate type.
The examples are
(c) the SF-Haldane transition in $S=1$ $XXZ$ spin chain, and
(d) the SF-dimer transition in $S=1/2$ bond-alternating $XXZ$ spin chain.
\cite{oka-sugi,yoshi-oka}
In the non-degenerate case, the gapped states
(the Haldane state in (c) and the dimer state in (d)) are unique and 
non-degenerate.
The nature of this type of transition is that the gap-formation
mechanism (for instance, the bond alternation in (d))
in the Hamiltonian is renormalized to zero in the sense of
the renormalization group treatment due to the strong quantum fluctuations.

In this paper, we review the {\it level spectroscopy}\cite{review,nomu-review}
(LS) by use of which
we can determine the BKT critical point very accurately
(typical accuracy is $10^{-3}$ or better)
from the numerical diagonalization data,
overcoming the above-mentioned difficulties.
We mainly focus on the physical meaning\cite{julien} of the LS for the doubly degenerate type
in the one-dimensional quantum spin models.
Also we explain how to apply the level spectroscopy to the
magnetization plateau problems.

\section{Physical Meaning of the Level Spectroscopy}

To distinguish the gapless spin-fluid state and the gapful state,
the most fundamental quantity is the excitation gap for infinite systems.
If we know the analytically exact solution,
we can easily distinguish the gapless and gapful states.
In usual cases, however, we do not know the exact solution,
so that we have to rely on the numerical methods,
for instance the numerical diagonalization.
Although we may extrapolate the gap data to $N\to\infty$ ($N$ is the system size),
the extrapolated data contains numerical errors,
which brings about the difficulty in judging whether the extrapolated value
is zero or finite.
Further, the extrapolation from the $N < \xi$ ($\xi$ is the correlation length)
data is unreliable.
If we use this type of extrapolation,
the apparent BKT critical point will be the point of $\xi \sim N_0$, 
where $N_0$ is the typical system size of the diagonalization data.
Thus this method results in the wider gapful region than the true one.

The shortcoming of the extrapolation method lies in
the zero-or-finite judgment.
If we find two quantities which cross with each other at the BKT critical point
as functions of quantum parameters in the Hamiltonian,
we can easily know the candidate for the critical point.
This is the fundamental idea of the LS.

\begin{figure}
    \epsfxsize= 10 cm   
    \centerline{\epsfbox{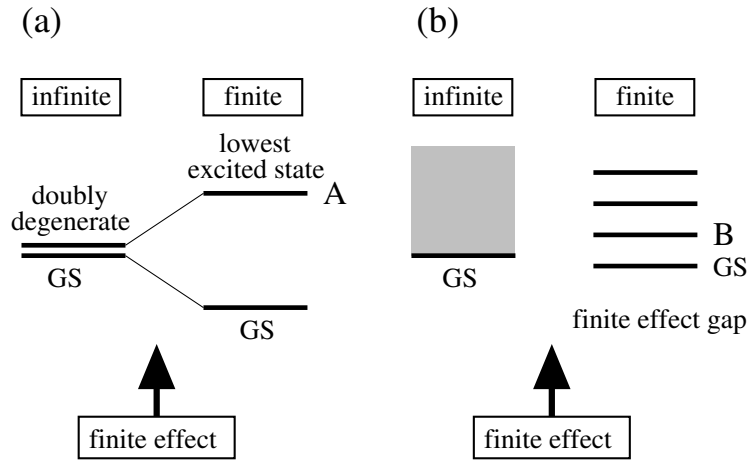}}
\caption{The finite effects are schematically shown for
         the (a) doubly degenerate ground state (gapless)
         and (b) the spin-fluid ground state (gapless).
         GS is the abbreviation of \lq\lq ground state\rq\rq.
         The shadowed region represents the continuous spectrum.
         }
\label{fig:1}
\end{figure}
In usual cases,
the ground state of the antiferromagnetic spin model is
unique (non-degenerate) for finite systems,
even when the degenerate ground state is realized for
infinite systems.
The exceptions are, for example, Ising model and the Majumdar-Ghosh model.
Then, how does the doubly degenerate ground state take place?
The mechanism is as follows.
A low-lying excitation of the finite system
asymptotically degenerate to the ground state as $N \to \infty$. 
In this limit, a recombination of these two states occurs, 
which results in the realization of the doubly degenerate ground states
with spontaneous symmetry breaking.
In other words,
the double degeneracy in infinite systems is lifted by the
perturbation of \lq finiteness\rq\ 
as is schematically shown in Fig. 1(a).
Then the lowest excitation in the gapful region
should be the broken half of the doubly degenerate states
denoted by A in Fig. 1(a). 
The property of the A state depends on what kind of doubly degenerate state
is realized in infinite systems.

On the other hand,
the gapless spectrum in the spin fluid case becomes discrete for finite systems.
The lowest excitation should be the spin-wave excitation
denoted by B in Fig. 1(b).

Then the properties of the lowest excitations for finite systems
in case of the SF and gapful cases
are quite different from each other.
Therefore the BKT critical point can be obtained
from the crossing of these two excitations
as functions of quantum parameters in the Hamiltonian.
Thus our method is named {\it level crossing method},
or more sophisticatedly {\it level spectroscopy}.

\section{An Example: ${\mib S=\mbox{\bf 1/2}}$ $\mib{XXZ}$ Spin Chain
 with Next-Nearest-Neighbor Interactions}

As an example, let us take up the $S=1/2$ $XXZ$ spin chain with
next-nearest-neighbor interactions\cite{tonegawa-harada,ON,NO1,NO2}
described by
\begin{equation}
    \cH
    = \sum_j \left\{ (\vS_j \cdot \vS_{j+1})_\Delta
                + \alpha (\vS_j \cdot \vS_{j+2})_\Delta \right\}
\end{equation}
where
\begin{equation}
    (\vS_l \cdot \vS_m)_\Delta
    \equiv S_l^x S_m^x + S_l^y S_m^y + \Delta S_l^z S_m^z
\end{equation}
\begin{wrapfigure}{r}{6.6cm}
  \epsfxsize= 5.5 cm   
  \centerline{\epsfbox{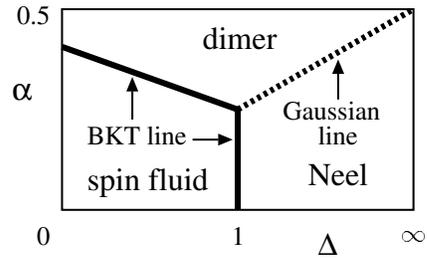}}
\caption{Phase diagram of the model (3.1).}
\label{fig:2}
\end{wrapfigure}
For simplicity we restrict ourselves to the
$0 \le \Delta < \infty$ and $0 \le \alpha \le 1/2$ case.
In this parameter region, the ground state phase diagram
consists of three phases, the SF phase, the dimer phase and the
N\'eel phase, as is schematically shown in Fig. 2.
The SF state is unique and gapless,
whereas the latter two states are doubly degenerate and gapful.

\begin{wrapfigure}{r}{6.6cm}
  \epsfxsize= 5.5 cm   
  \centerline{\epsfbox{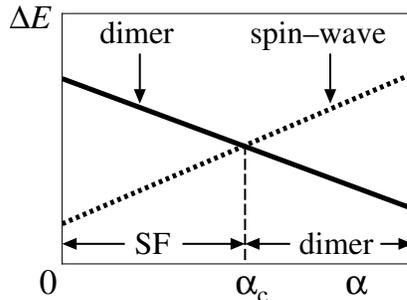}}
\caption{Schematic behavior of the dimer excitation with $S_{\rm tot}=0$
         (solid line) and the spin-wave excitation with $S_{\rm tot}=1$
         (dotted line).
         The ground state should be the dimer state or the SF state
         depending on whether the former is smaller or larger then the latter.}
\label{fig:cross}
\end{wrapfigure}
We consider the $\Delta=1$ case where the total spin
$S_{\rm tot}$ is a good quantum number, for example.
When $N=4n$ case ($n$ is an integer),
the ground state for the finite systems is of $S_{\rm tot} = 0$.
Since the dimer ground state is also of $S_{\rm tot}=0$,
the lowest excited state in the dimer region,
which is nothing but the broken half of the doubly degenerate
ground states for infinite systems,
should be also of $S_{\rm tot}=0$
due to the addition rule of the angular momentum.
On the other hand, 
the lowest excitation in the SF region should be of the spin-wave type
with $S_{\rm tot} =1$ (triplet due to the $SU(2)$ symmetry).
Thus the crossing of the $S_{\rm tot}=0$ and $S_{\rm tot}=1$
excitations is the candidate for the SF-dimer critical point,
as schematically shown in Fig. 3.

Since the crossing point is weakly dependent on $N$,
the final result is known by taking $\lim_{N\to\infty} \alpha_{\rm c}(N)$.
By this procedure, we can obtain $\alpha_{\rm c} = 0.2411$ for
$\Delta=1$.
We note that the only 4 spin result is
$\alpha_{\rm c}(N=4) = 0.25$,
which suggests the effectiveness of the LS method.

\begin{wraptable}{r}{\halftext}
\caption{Symmetry classification of the ground state and important excitations.}
\label{table:1}
\begin{center}
\begin{tabular}{ccccc} \hline \hline
excitation &$S_{\rm tot}^z$ & $k$  & $P$ &$T$ \\ \hline
ground state     & 0  & 0 &$+1$ &$+1$ \\
dimer excitation     & 0  & 0  &$+1$  &$+1$ \\
N\'eel excitation     & 0  & $\pi$ &$-1$ & $-1$ \\ 
spin-wave excitation     & $\pm 1$  & $\pi$ &$-1$ &$*$\\ 
\hline
\end{tabular}
\end{center}
\end{wraptable}
When $\Delta \ne 1$,
$S_{\rm tot}$ is not a good quantum number.
In this case,
we should classify the excitations by discrete eigenvalues of the symmetry
operators.
They are $S_{\rm tot}^z$, the wave number $k$,
the eigenvalues of the parity operator $P$ ($S_j \leftrightarrow S_{N-j+1}$)
and the time reversal operator
($|\uparrow\rangle \leftrightarrow |\downarrow\rangle)$.
Similar consideration to the $\Delta=1$ case leads to
the classification of the excitations as shown in Table 1.
The $T$ operator is valid for the $S_{\rm tot}^z=0$ case.
We can also find the dimer-N\'eel transition point
from the crossing of the dimer and N\'eel excitations.
This dimer-N\'eel transition is of the second order with the
Gaussian universality in which the critical exponents varies
continuously along the transition line.

\section{Application to the Magnetization Plateau Problems}

Since the magnetization plateau is related to the field-induced
excitation gap,
we can apply the LS method to this problem.
There are also two types of mechanisms in the magnetization plateau,
as in case of the zero magnetization case.
The examples of the doubly degenerate type are
(a) $M_{\rm S}/2$ plateau in the $S=1/2$ Heisenberg chain with
bond-alternation and the next-nearest-neighbor interactions,\cite{tone1,tone2} 
and
(b) $(2/3)M_{\rm S}$ plateau in the $S=1/2$
distorted diamond chain.\cite{tone-oka,tone-oka2}
For the non-degenerate case, the examples are
(c) $M_{\rm S}/3$ plateau in the $S=1/2$
ferromagnetic-ferromagnetic-antiferromagnetic spin chain,\cite{FFA1,FFA2,FFA3}
and
(d) $M_{\rm S}/$ plateau in the $S=3/2$ $XXZ$ spin chain with
on-site anisotropy.\cite{kita-oka,oka-kita}
We note that the model (c) is the first model in which
the magnetization plateau in quantum spin chain is discussed.

Because we should compare the lowest energies $E_0$ with the magnetization
$M=M_0,M_0 \pm 1$ 
in the plateau problems at the magnetization $M_0$,
a slight modification from the no magnetic field case is necessary.
The difference in $M$ corresponds to the difference in the
number of fermions in the fermion picture.
Thus we have to take the effect of the chemical potential $\mu$
into consideration.
Namely, we should compare $E_0(M_0)$ and $E_0(M_0+1)-\mu$,
and also $E_0(M_0)$ and $E_0(M_0-1)+\mu$.
Since the chemical potential around $M_0$ is given by
\begin{equation}
    2\mu
    = E_0(M_0 + 1) - E_0(M_0 - 1)
\end{equation}
we see
\begin{eqnarray}
    &&\{E_0(M_0+1)-\mu\} - E_0(M_0)
    = \{E_0(M_0-1)+\mu\} - E_0(M_0)   \nonumber \\
    &&= (1/2)\{E_0(M_0+1) + E_0(M_0-1) \} - E_0(M_0)
\end{eqnarray}
which plays the role of the spin-wave excitation in the
zero magnetic field case.
The same expression can be obtained by use of the
Legendre transformation $E \to E-MH$.

Using this method,
we have obtained the plateau phase diagrams of
the $(2/3)M_{\rm S}$ plateau of the $S=1/2$
distorted diamond spin chain,\cite{tone-oka,tone-oka2}
the $M_{\rm S}/2$ plateau of the $S=1/2$
frustrated two-leg spin ladder,\cite{okazaki}
the $M_{\rm S}/2$ plateau of the $S=1/2$
two-leg spin ladder with 4-spin cyclic exchange interactions,\cite{nakasu}
the $M_{\rm S}/4$ plateau of the $S=1$
frustrated two-leg spin ladder.\cite{oka-okazaki1,okazaki2,oka-okazaki2} 
Yamamoto, Asano and Ishii\cite{YAI}
applied this method to the Kondo necklace model with next-nearest-neighbor
interactions in a successful way.

\section{Summary}

We have explained the level spectroscopy (LS) especially focusing on its physical
meaning in the doubly degenerate cases.
Not only we can obtain the BKT critical point from the level crossing,
but also we can check the universality class
by the careful examination of the combination of the low-lying excitations
to eliminate the troublesome logarithmic size corrections
in the lowest order.\cite{NO2,nomura}
Thus the name {\it level spectroscopy} may be more appropriate than the
name {\it level crossing method} for our method.

For the non-degenerate case,
of which we have mentioned only briefly,
the LS method is also established,\cite{kitazawa,kita-nomu,kita-nomu-oka,kita-nomu2,kita-nomu3}
although it is more complicated and the
physical meaning is not so clear as the degenerate case.
The use of the twisted boundary condition\cite{kitazawa,kita-nomu}
has been proved to be powerful in many cases
of the zero-magnetic field BKT transitions
as well as the plateauful-plateauless transitions.
\cite{FFA3,kita-oka,oka-kita,oka-kita2,oka-kita3,chen1,chen2}
This twist boundary condition method is also applicable to the 
transition between two plateau states with different plateau 
formation mechanisms.\cite{kita-oka,oka-tone,sakai-oka}

The LS method has been applied not only to the quantum spin models 
but also to the lattice electron  models, for instance,
 $t-J$ and related models,\cite{nakamura1,nakamura1aa,nakamura1a}
and the extended Hubbard and related
models.\cite{nakamura1a,nakamura2,nakamura3,otsuka1,otsuka2,nakamura4,nakamura5}

\section*{Acknowledgements}
We would like to express our appreciation to Kiyohide Nomura, Atsuhiro Kitazawa,
Takashi Tonegawa and T\^oru Sakai for collaborations in
the developments and applications of the level spectroscopy.
We also thank Masaaki Nakamura for stimulating discussions.

\end{document}